\documentclass[aps,prb,twocolumn]{revtex4-1}
\usepackage{amsmath}
\usepackage{graphicx}
\usepackage{amsfonts}
\usepackage{amssymb}
\usepackage{color}
\usepackage{enumitem}
\usepackage{multirow}

\begin{document}
\title{Band twisting and resilience to disorder in long-range topological superconductors} 

\author{T. O. Puel}
\email{tharnier@csrc.ac.cn}
\affiliation{Beijing Computational Science Research Center, Beijing 100193, China}
\affiliation{CeFEMA, Instituto Superior T\'ecnico, Universidade de Lisboa Av. Rovisco
Pais, 1049-001 Lisboa, Portugal}
\affiliation{Zhejiang Institute of Modern Physics, Zhejiang University, Hangzhou,
Zhejiang 310027, China}

\author{O. Viyuela}
\email{oviyuela@fas.harvard.edu}
\affiliation{Department of Physics, Harvard University, Cambridge, MA 02318, USA}
\affiliation{Department of Physics, Massachusetts Institute of Technology, Cambridge,
MA 02139, USA}

\date{\today}

\begin{abstract}
Planar topological superconductors with power-law-decaying pairing display different kinds of topological phase transitions where quasiparticles dubbed nonlocal-massive Dirac fermions emerge. These exotic particles form through long-range interactions between distant Majorana modes at the boundary of the system. We show how these propagating-massive Dirac fermions neither mix with bulk states nor Anderson-localize up to large amounts of static disorder despite being finite energy. Analyzing the density of states (DOS) and the band spectrum of the long-range topological superconductor, we identify the formation of an edge gap and a surprising double peak structure in the DOS which can be linked to a twisting of energy bands with nontrivial topology. Our findings are amenable to experimental verification in the near future using atom arrays on conventional superconductors, planar Josephson junctions on two-dimensional electron gases, and Floquet driving of topological superconductors.
\end{abstract}

\maketitle

\section{Introduction}

Symmetry-protected topological (SPT) orders are quantum phases of matter characterized by nonlocal order parameters (topological invariants) and protected edge states at the boundary \cite{rmp1,rmp2}. SPT phases with particle-hole symmetry give rise to topological superconductors \cite{Read_et_al00,LibroBernevig} with unconventional pairing and gapless edge states, dubbed Majorana zero modes (MZMs). MZMs are nonabelian anyons, which can be braided to perform topological quantum computation and are protected against thermal fluctuations by a superconducting gap \cite{rmp3,rmp4,Alicea_et_al_11,Baranov_et_al_13,Mazza_et_al13}. These unpaired Majorana particles were first shown to arise at the ends of a chain of fermions with $p$-wave superconducting pairing \cite{Kitaev01}. However, the impracticality of $p$-wave pairing in nature was initially believed to be a roadblock, until proximity induced superconductivity schemes have proven to be a way to circumvent this obstacle \cite{Fu_Kane_2008}. 

In recent years, different experiments have shown Majorana physics by means of a conventional superconductor proximitized to the surface of a topological insulator \cite{Fu_Kane_2008,bib:Xu2015,bib:Sun2016}, semiconductor nanowires with strong spin-orbit coupling and subject to Zeeman fields \cite{Sau_et_al_2010, Alicea_2010,Sau2_et_al_2010,Sau3_et_al_2010,Oreg_et_al_2010,Mourik_et_al12,Deng_et_al12,Das_et_al12,Wang_et_al12,bib:Rokhinson2012,bib:Deng2013,He_et_al14,Albrecht_et_al16,bib:Sun2016,bib:Deng2016,He_et_al17,bib:Lutchyn2018}, quantum anomalous Hall insulator-superconductor structures \cite{He_et_al17}, and atomic arrays on superconducting substrates \cite{Pientka2013,bib:Braunecker2013,Nadj_et_al13,Klinovaja_et_al13,Pientka2014,Li_et_al_16,Kaladzhyan_et_al16,Kaladzhyan_et_al16B,Nadj_et_al14,bib:Ruby2015,Pawlak_et_al16,Ruby_et_al17,Menard_et_al15,Menard_et_al17,Heinrich_et_al17,Ronty_et_al_15,Li_et_al_16_2D}. In particular, one-dimensional arrays of magnetic impurities \cite{Pascual_et_al16,Ruby_et_al17}, where the length of the chain is relatively small compared to the coherence length of the host superconductor \cite{Pientka2013}, generates an effective $p$-wave Hamiltonian with long-range pairing \cite{Nadj_et_al13,Pientka2013,Klinovaja_et_al13,Pientka2014,Li_et_al_16,Kaladzhyan_et_al16,Kaladzhyan_et_al16B}. Floquet driving a $p$-wave superconductor \cite{Benito_et_al14} and planar Josephson junctions proximitized to a 2D electron gas (2DEG) with spin-orbit coupling and Zeeman field \cite{PhysRevX.7.021032,Liu_et_al18,Fornieri:2019aa} also give rise to effective models of topological superconductivity with long-range couplings. 

Inspired by these recent experimental developments, $p$-wave Hamiltonians with long-range couplings have been throughly studied \cite{Niui_12,DeGottardi_13,Vodola_et_al14,Tudela_15,Vodola_et_al16,Viyuela_et_al16,Gong2016_1,Gong2016_2,Pachos_17,Lepori_17,Alecce_17,Vodola_et_al17,Dutta_17,Cats_et_al18,Giuliano_18,Viyuela_et_al18,Lepori_18}. A long-range extension of the Kitaev chain with power-law-decaying hopping and pairing amplitudes give rise to a combined exponential and algebraic decay of correlations, breakdown of conformal symmetry and violation of the area law of entropy \cite{Vodola_et_al14,Vodola_et_al16}. The topological nature of this new model has been also unveiled \cite{Viyuela_et_al16}, demonstrating the existence of fractional topological numbers associated to nonlocal-massive Dirac fermions \cite{Viyuela_et_al16,Lepori_17,Alecce_17}. These particles are fermions with a highly nonlocal extension, as they are formed out of the long-range interaction of distant Majorana particles at the edge, and their localization properties are indeed robust to weak static disorder \cite{Viyuela_et_al16}. Interestingly, a staircase of higher-order topological phase transitions can be induced by tuning the exponent of the power-law-decaying pairing amplitude \cite{Cats_et_al18}.

Generalizations of the long-range Kitaev chain to two-dimensions have been constructed \cite{Viyuela_et_al18,Lepori_18}, where the $p$-wave character of the superconductor is preserved while including power-law-decaying couplings that extend over the plane. In these systems, topological phases holding propagating Majorana edge states with different chiralities get significantly enhanced by long-range couplings. In one of the topological phases, propagating Majorana fermions at each edge pair nonlocally and become gapped for sufficiently long-range interactions, while remaining topological and localized at the boundary \cite{Viyuela_et_al18}. However, the robustness of these new chiral edge states with respect to general static disorder was unclear and the effects of the long-range couplings in the band spectrum of the topological superconductor were not explored. 


In this article, we study how propagating Majorana states, which become gapped by the effect of long-range interactions, are affected by the inclusion of static disorder. We show how the localization at the edge is preserved even for very strong disorder, demonstrating that the propagating massive Dirac fermions at the edge are not pushed to the bulk nor get delocalized. This is one of the characteristic features of all topologically protected edge states. 
Moreover, we study how the band spectrum of a planar $p$-wave topological superconductor is modified by the effect of long-range couplings. We prove how a characteristic (and previously unnoticed) double peak structure in the density of states (DOS) of the topological superconductor is enhanced by the inclusion of power-law-decaying amplitudes. Associated with that effect we find a band twisting in the energy spectrum provided the phase is topologically nontrivial.


The paper is structured as follows. In Sec. \ref{sec:II}, we introduce the 2D $p$-wave Hamiltonian with long-range couplings and perform a detailed study of the band structure and the density of states as function of the decay exponents. In Sec. \ref{sec:III} we demonstrate the robustness of the nonlocal-massive Dirac fermions due to disorder and compare it to the case with unpaired Majoranas through the spatial distribution of those nonlocal-massive Dirac fermions. Sec. \ref{sec:IV} is devoted to conclusions. 
In the Appendix we perform a finite size scaling of ingap states and their dependance on the decay exponent $\alpha$, 
and analyze the robustness of the system with respect to different types of static disorder.

\section{Band Structure $\&$ Density of States}
\label{sec:II}

The model studied in this paper is that of a two-dimensional spinless $p$-wave superconductor with
long-range hopping and long-range superconducting coupling. In real space the Hamiltonian can be written as
\begin{align}
H & =-\left(\mu-4t\right)\sum_{\boldsymbol{r}=1}^{N}\left(c_{\boldsymbol{r}}^{\dagger}c_{\boldsymbol{r}}-c_{\boldsymbol{r}}c_{\boldsymbol{r}}^{\dagger}\right)\nonumber \\
 & -\sum_{\boldsymbol{r}}\sum_{\boldsymbol{r}'\neq\boldsymbol{r}}\frac{t}{R^{\beta}}\left(c_{\boldsymbol{r}'}^{\dagger}c_{\boldsymbol{r}}+c_{\boldsymbol{r}}^{\dagger}c_{\boldsymbol{r}'}\right)\nonumber \\
 & -\sum_{\boldsymbol{r}}\sum_{\boldsymbol{r}'\neq\boldsymbol{r}}\frac{\Delta}{R^{\alpha+1}}\left[\left(R_{x}+iR_{y}\right)c_{\boldsymbol{r}'}^{\dagger}c_{\boldsymbol{r}}^{\dagger}+\left(R_{x}-iR_{y}\right)c_{\boldsymbol{r}}c_{\boldsymbol{r}'}\right],\nonumber \\
\label{eq:Hamiltonian real space}
\end{align}
where both $\boldsymbol{r}$ and $\boldsymbol{r}'$ run over 
all sites of a square lattice
labelled from $1$ to $N$, where $N$ is the total number of
sites. We have defined $\boldsymbol{R}=\left(R_{x},R_{y}\right)\equiv\boldsymbol{r}-\boldsymbol{r}'$
and $\left|\boldsymbol{R}\right|=\sqrt{R_{x}^{2}+R_{y}^{2}}\equiv R$.
The band width is represented by $t$ and the coupling strength is represented by $\Delta$.
The exponents $\alpha$ and $\beta$ control the decay of superconducting
coupling range and hopping range, respectively. The chemical potential
$\mu$ eventually drives the system to phase transitions, for example
in the regime of fast decay (large values of the decaying exponents)
we find a transition from a trivial superconducting phase
(SC) to a topological superconducting phase characterized by Majorana fermions ($\cal M$).
Interestingly, it is known that long-range superconducting couplings
give rise to new topological phases characterized by massive Dirac fermions
($\cal D$). 
This phase transition happens at the critical value $\alpha=2$ and only exists for one of the two topological phases \cite{Viyuela_et_al18}.
This differs from the semi-2D Hamiltonian\cite{Lepori_18}, where the long-range terms appear only in $x$ and $y$ directions. 
The phase transition then occurs at $\alpha = 1$ and is present in both topological phases.
A phase diagram illustrating the former case is depicted in Fig.\ref{fig1}a.
Unless explicitly mentioned, we have used $t=0.5$ as reference parameter, $\Delta=0.5$
following Ref. [\onlinecite{Viyuela_et_al18}], and $\beta=10$, 
i.e. fast-decay hopping. 
For instance we have verified that whatever $\beta, \alpha  \geq 20$ gives the same energy spectrum of the pure short-range hopping, with next-nearest neighbors hopping.

\subsection{Massive Dirac fermions}

The first step is to identify the differences and similarities between the Majorana phase and the massive Dirac phase. For that, the edge-state excitations will be analysed.

\begin{figure}
\includegraphics[width=0.65\columnwidth]{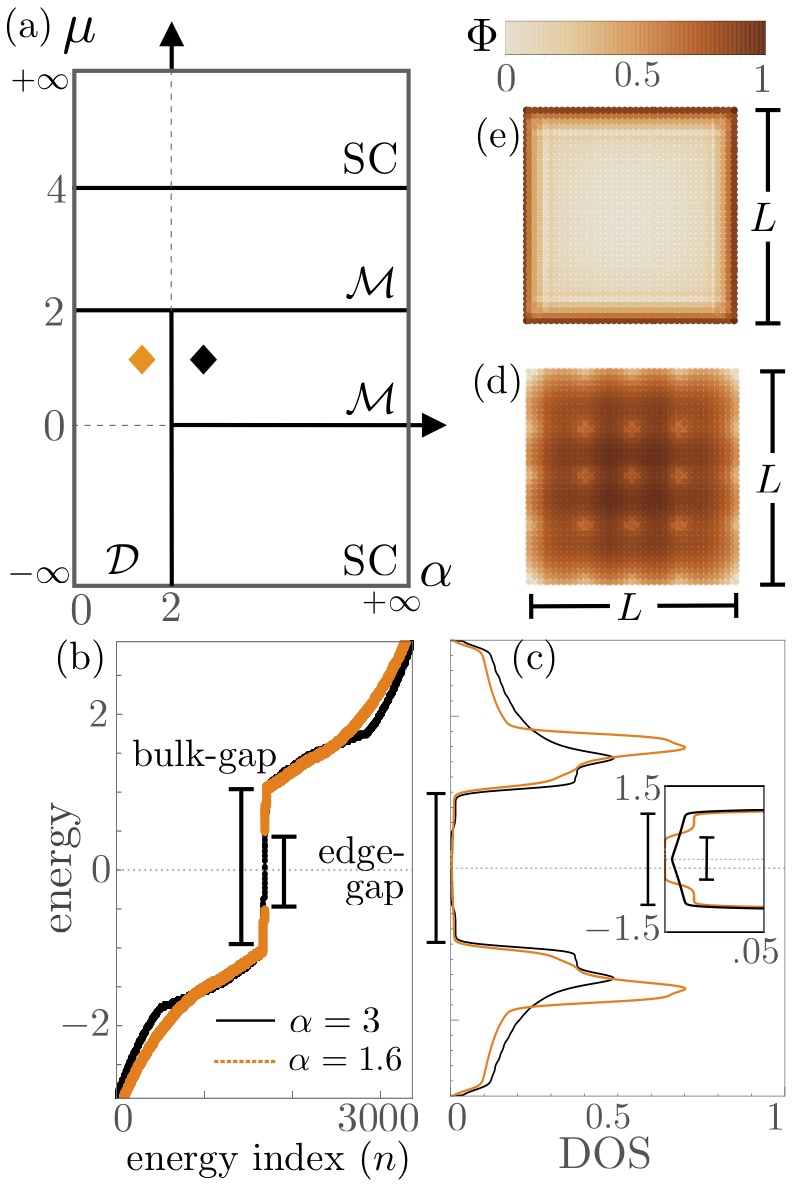}

\caption{Panel (a) presents the phase diagram for a range of chemical potential,
$\mu$, and long-range superconducting coupling, $\alpha$, parameters.
Three different phases can be identified: 
(i) a trivial superconducting phase, SC;
(ii) a topological superconducting phase with Majorana fermions, $\cal M$; 
(iii) a topological superconducting phase with massive Dirac fermions, $\cal D$. 
We note that the two different phases $\cal M$ have opposite chiralities.
Plots (b) and (c) show the energy spectrum and the DOS, respectively, for the two topological phases (signalled in the phase diagram, using the same color code) 
in a system of size $N = 1681$.
Panels (d) and (e) show the probability
of occupancy associated to the $n$-th energy of the 2D finite-squared system (top view) in the $\cal D$ phase,
as described in the the previous panels. 
In particular, a representative probability of occupancy for a bulk energy is plotted in panel (d), while the lowest finite energy inside the gap is plotted in panel (e).
Note that the probability of occupancy is plotted in log scale, thus written in terms of $\Phi$ as defined in the main text.}
\label{fig1}

\end{figure}

By exactly diagonalization $H\left|\psi_{n}\right\rangle =E_{n}\left|\psi_{n}\right\rangle $
we obtained the Bogoliubov energy spectrum, $E_{n}$ with $n=1,\ldots,2N$,
of a finite (squared) system with $L^{2}\equiv N$ lattice sites.
The results are depicted in Fig.\ref{fig1}, in which we exemplified the two different topological phases $\cal M$ and $\cal D$.
The parameters are indicated in the phase diagram, panel (a), by the geometric figures in diamond shape, namely we set $\alpha=1.6$ and $\alpha=3$, with $\mu=1$.
In both phases, the superconducting gap (stated here as bulk-gap) is easily noticed from either the energy spectrum in panel (b) or its respective density of states (DOS) in panel (c). 
The topological properties are manifested as ingap states, in particular the inset of panel (c) explicits the difference between the two topological 
phases\footnote{See Appendix \ref{app: finite size scaling} for a finite size scaling analysis of the ingap states and their dependance on the decay exponent $\alpha$.}.
While the Majorana states manifest as a finite DOS over the entire gap, the massive Dirac states let opened a smaller gap (stated here as edge-gap since it is the energy difference between edge-state excitations).

One may also look at the localization of massive Dirac states plotting
the probability of occupancy related to the $n$-th wave-vector (corresponding
to energy $E_{n}$ inside the bulk-gap) on each site, i.e. ${\cal P}_{n}\left(\boldsymbol{r}\right)\equiv a_{n}\left(\boldsymbol{r}\right)a_{n}^{*}\left(\boldsymbol{r}\right)$,
where the amplitude $a_{n}\left(\boldsymbol{r}\right)$ is obtained
from $\left|\psi_{n}\right\rangle =\sum_{\boldsymbol{r}}a_{n}\left(\boldsymbol{r}\right)\left|\psi_{n}\left(\boldsymbol{r}\right)\right\rangle $,
and the normalization implies $\sum_{\boldsymbol{r}}{\cal P}_{n}\left(\boldsymbol{r}\right)=1$.
Figs.\ref{fig1}d and \ref{fig1}e exemplify this probability for an energy inside the bulk and for the smallest finite energy inside the bulk-gap, respectively.
The probability amplitude of occupancy is better analysed if log scaled, thus for convenience we have defined a normalized logarithmic localization $\Phi \equiv 1 - \log{\cal P}_{n}\left(\boldsymbol{r}\right) / \log {\cal P}_{\text{min}}$, 
where $\Phi = 1$ if ${\cal P}_{n}\left(\boldsymbol{r}\right) = 1$ and $\Phi = 0$ if ${\cal P}_{n}\left(\boldsymbol{r}\right) = {\cal P}_{\text{min}}$.
${\cal P}_{\text{min}}$ is the global-minimum probability ${\cal P}_{n}\left(\boldsymbol{r}\right)$, i.e. among all energies $E_n$ and all sites $\boldsymbol{r}$.

Equivalent to the Majorana excitations in the planar topological superconductor, the massive Dirac states are confined to the edges, see Fig.\ref{fig1}e, which form propagating modes protected by particle-hole symmetry. 
Technically speaking, the system still belongs to class D of topological superconductors~\cite{Schnyder2009} with $\mathbb Z$ topological invariant
\footnote{For instance, in $k$-space the Hamiltonian assumes the form $H = \text{even($k$)}\sigma_z + \text{odd($k$)} (\sigma_x + i \sigma_y)$, where $\sigma$ acts on the Nambu basis. Thus, the particle-hole operator is ${\cal P} \equiv \sigma_x K$, which satisfies the relation $H_k = - {\cal P} H_{-k} {\cal P}$, or the relation $H = - {\cal P} H^T {\cal P}$ in real space.  It also has inversion symmetry, whose operator is ${\cal I} \equiv \sigma_z$ and respects the relation $H_k = {\cal I} H_{-k} {\cal I}$, or the relation  $H_{\boldsymbol{R}} = {\cal I} H_{-\boldsymbol{R}} {\cal I}$ in real space.}.
In Fig.\ref{fig1}d we see the bulk energy excitations remaining spread over the sample.
A thorough study of the robustness of the massive Dirac states is one of the main goals of this work and will be discussed in section \ref{sec:Localization-and-Robustness}.

\subsection{Twisted bands and double peak structure}

We discovered that the band spectrum and the DOS of our long-range topological superconductor provide valuable informations regarding the energy distribution of the different eigenstates (see Fig.\ref{fig1}c). In addition, we may extract useful quantities such as the magnitude of the superconducting gap, the group velocity and the band dispersion.

\begin{figure}
\includegraphics[width=0.95\columnwidth]{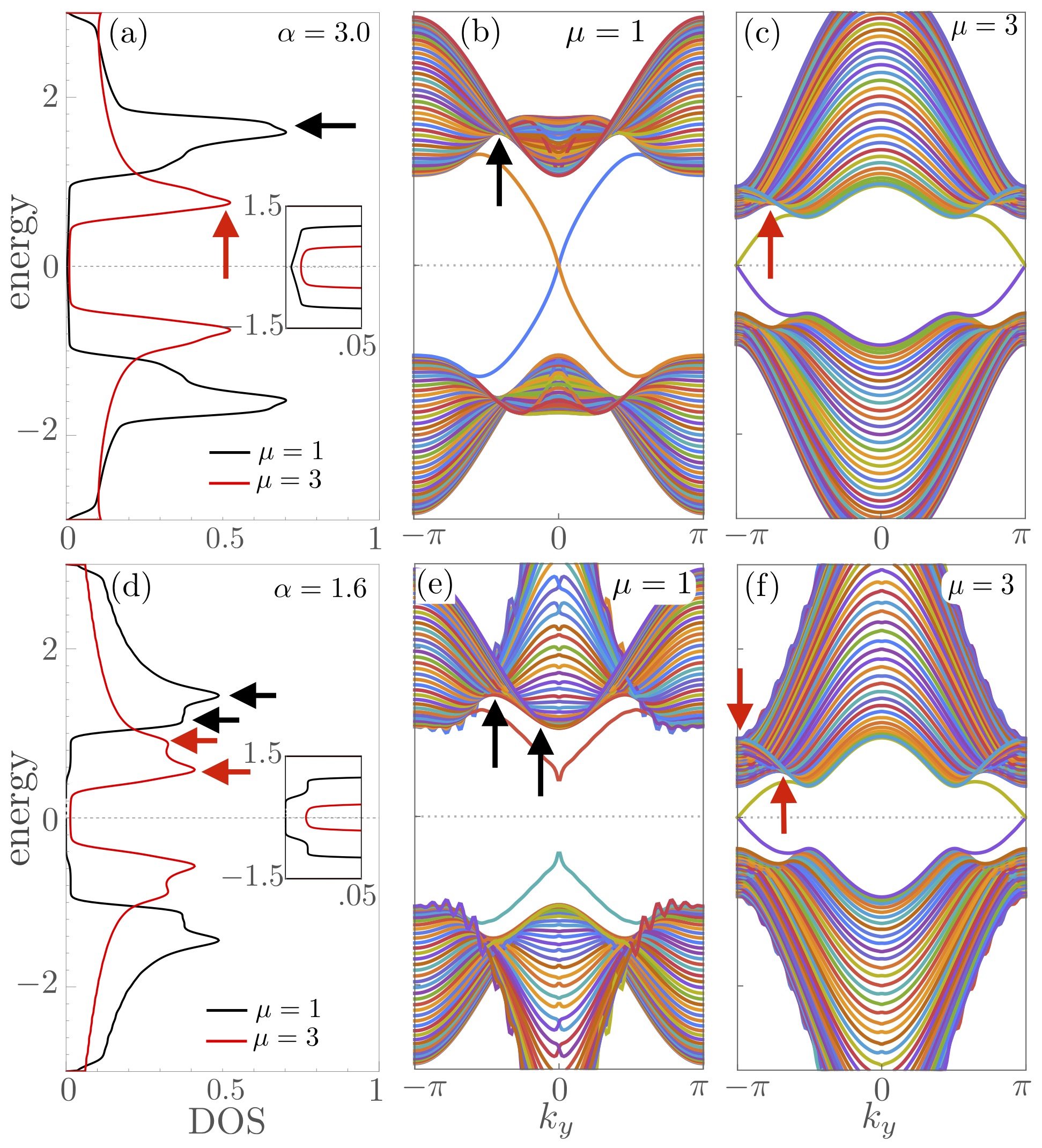}

\caption{Panel (a) shows the DOS of a finite squared system for the two phases $\cal M$ with different chiralities, namely $\mu = 1$ and $\mu = 3$,
and the inset is a zoom to the ingap states.
Panels (b) and (c),  respectively, show the corresponding band spectrum for a semi-infinite system, i.e. periodic in the $y$ direction.
The many different colors represent different energy levels.
The arrows indicate the two peaks structure on the DOS, and their associated band twist in the band spectrum.
Panels (d)-(f) show equivalent results for longer range couplings,
in particular note that for $\mu=1$ the system is in the phase $\cal D$.}

\label{fig2}
\end{figure}

For convenience, we consider a semi-infinite system, finite in $x$ direction and periodic in the $y$ direction. 
As an example, let's take two points in the phases $\cal M$ of the phase diagram with different chiral edge states, namely $\mu=1$ and $\mu=3$, with $\alpha=3$.
Fig.\ref{fig2}a shows the DOS of these two points, while Figs.\ref{fig2}b
and \ref{fig2}c show their respective band spectrum for a semi-infinite system. 
From these figures we highlight the following: (i) associated to the peak
structures we notice an unusual band twisting (highlighted by the
arrows), and (b) there is a significative bands overlap as consequence
of this band twisting. 

The double peak structure in the DOS is a measurable consequence of band inversion in topological superconductors.
For instance, if the two particle-hole symmetric bands overlapping for small values of $\Delta$, as we enlarge the superconducting amplitude a gap is opened and a band inversion is formed. 
Such a band inversion does not happen in the trivial phases.
Most notably, in the long-range system with slow-decaying coupling strength, the band twist (or band inversion) occurs even when the particle-hole bands do not overlap in the limit $\Delta \rightarrow 0$.
This behavior leads to a higher concentration of density of states around two areas where the twisting of bands occurs, which in turn generates a double peak structure in the DOS.

Next we observe that longer range superconducting couplings
are responsible for the enhancement of the peak's structure, in particular within the massive Dirac phase $\cal D$. 
Figs.\ref{fig2}d-f show the results for smaller values of the superconducting coupling exponent already inside the phase $\cal D$, i.e.
$\alpha=1.6$. We clearly see a more pronounced structure of the peaks,
more precisely they split into two peaks that comes along with an enlargement of the bands overlap.
We further note that the two peaks structure is present
in both topological phases, and that it is enhanced by decreasing
$\alpha$, however they do not appear in the trivial superconducting phase (not shown in this figure).

\begin{figure}
\includegraphics[width=0.95\columnwidth]{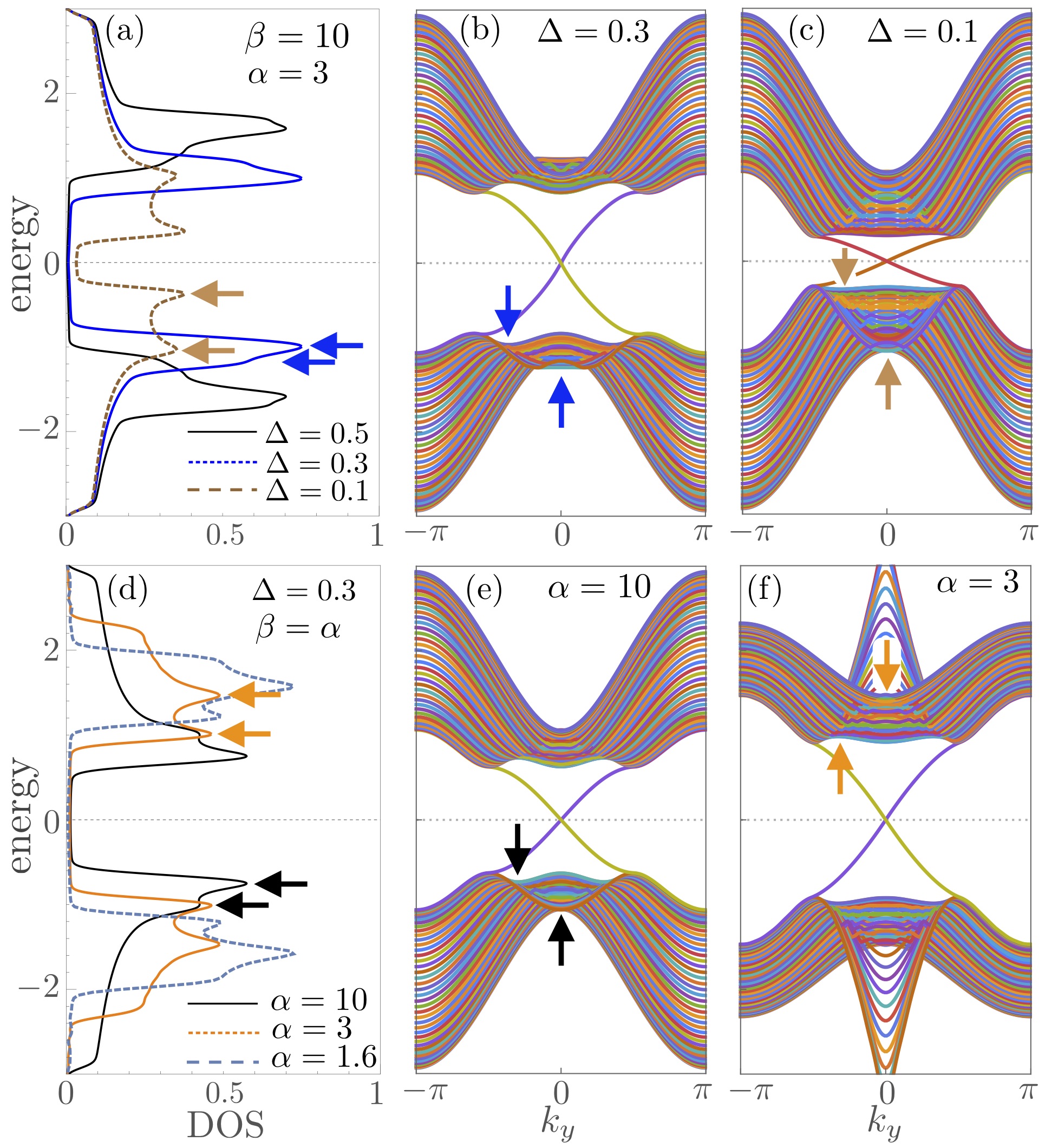}

\caption{Here we present analogous results as in Fig.\ref{fig2}.
Panels (a)-(c) show different values of the superconducting coupling strength $\Delta$, inside the phase $\cal M$ with $\mu = 1$.
Panels (d)-(f) show different values of both the superconducting coupling range ($\alpha$) and the hopping range ($\beta$), for $\mu=1$ and $\Delta = 0.3$. 
Note that the change of $\beta$ is not represented in the phase diagram of Fig.\ref{fig1}a, but for all the parameter's values shown here the system remains in the phase $\cal M$.}

\label{fig3}
\end{figure}

The superconducting coupling strength is also responsible for changing
the peak structure. In particular, decreasing $\Delta$ also makes
the peak split into two, as shown in Fig.\ref{fig3}a. Associated
with that, from the semi-infinite system band spectrum shown in Figs.\ref{fig3}b
and \ref{fig3}c, we again notice an enlargement of the bands overlap.
Indeed, we checked that lowering $\Delta$ (but finite) the two peaks
structure can always be retrieved in all topological phases.

The two peaks structure is not a unique long-range feature. In Figs.\ref{fig3}d-f
we show the presence of the two peaks even in the 
fast-decaying limit ($\left(\beta,\alpha\right)\gg1$).
And we have verified that these results match those from a system with short-range hopping.
In short, both topological phases present in this work ($\cal M$ and $\cal D$) present a double peak
structure in their DOS which is associated to a band twisting, which
in turn leads to a band overlap. 
This association is highlighted by the colored arrows in Figs. \ref{fig2} and \ref{fig3}.
Surprisingly the DOS double peak structure only appears within the topological phases. 
It is always achieved for finite-small values of the superconducting coupling strength and is enhanced by
longer-range couplings.

Therefore, within the limitations of the present model these double peak structures witness nontrivial band topology, due to the effect of band twisting. These results may help us distinguish more easily between different topologically trivial and nontrivial phases in experiments.

\subsection{Physical relevance of long-range couplings}

As already mentioned in the introduction, $p$-wave superconductors with long-range couplings naturally appear in different experimental realizations of these materials. 
A 2D sublattice of magnetic impurities, deposit on the surface of a conventional superconductor, leads to effective long-range pairing and hopping terms with a $1/\sqrt{r}$ decay \cite{Menard_et_al17}.  In particular, Mn adatoms deposit on top of Pb (001) have been shown to present long-range oscillations of up to 7-8 nm \cite{Heinrich_et_al17}, which proves the relevance of long-range interactions in these experiments.
We can also consider a different construction, where proximitizing planar Josephson junction to a 2DEG with Rashba spin-orbit coupling and Zeeman field produces an effective one-dimensional (1D) Kitaev chain with long-range pairing and hopping terms \cite{PhysRevX.7.021032,Liu_et_al18,Fornieri:2019aa}. The couplings of the effective 1D system can be tuned by varying the superconducting phase difference of the junction $\phi$, the inplane magnetic field $B$ and the chemical potential $\mu$. The emerging long-range couplings can be intuitively understood as arising from integrating out closely spaced modes residing along the transverse direction of the 2DEG. A similar construction could be used so that the integration of a 3D structure leads to effective long-range couplings in 2D. 
Finally, periodically driving a short-range topological insulator produces interesting effective models of 1D $p$-wave superconductors where long-range superconductivity arise \cite{Benito_et_al14}. Analogously, Floquet driving a planar $p$-wave superconductor would allow the tuning of effective long-range couplings. 
In conclusion, we have identified several experimentally relevant situations where the inclusion of long-range coupling terms is needed and where the physics of the topological superconductors described in this paper can be potentially tested.


\section{Robustness of the massive edge states against disorder\label{sec:Localization-and-Robustness}}
\label{sec:III}

Here we discuss the effect of static disorder in the presence of massive
Dirac states. We first analyse the normalized DOS computed
for a finite 2D system with different disorder strengths. The disorder
is added to the Hamiltonian as
\begin{equation}
H_{\text{disorder}}=\nu\sum_{\boldsymbol{r}=1}^{N}D_{\boldsymbol{r}}\left(c_{\boldsymbol{r}}^{\dagger}c_{\boldsymbol{r}}-c_{\boldsymbol{r}}c_{\boldsymbol{r}}^{\dagger}\right),
\end{equation}
where $\nu$ is the disorder strength and 
$\left|D_{\boldsymbol{r}}\right|\le1$ is equally distributed over the sites' positions $\boldsymbol{r}$.
Other realistic disorder distributions, such as a Gaussian peaked at $\mu$, would be less detrimental to our system and would serve as a less effective test of robustness for the edge 
states\footnote{See Appendix \ref{app: Gaussian disorder} where we compare different types of disorder.}.

\begin{figure}
\includegraphics[width=0.95\columnwidth]{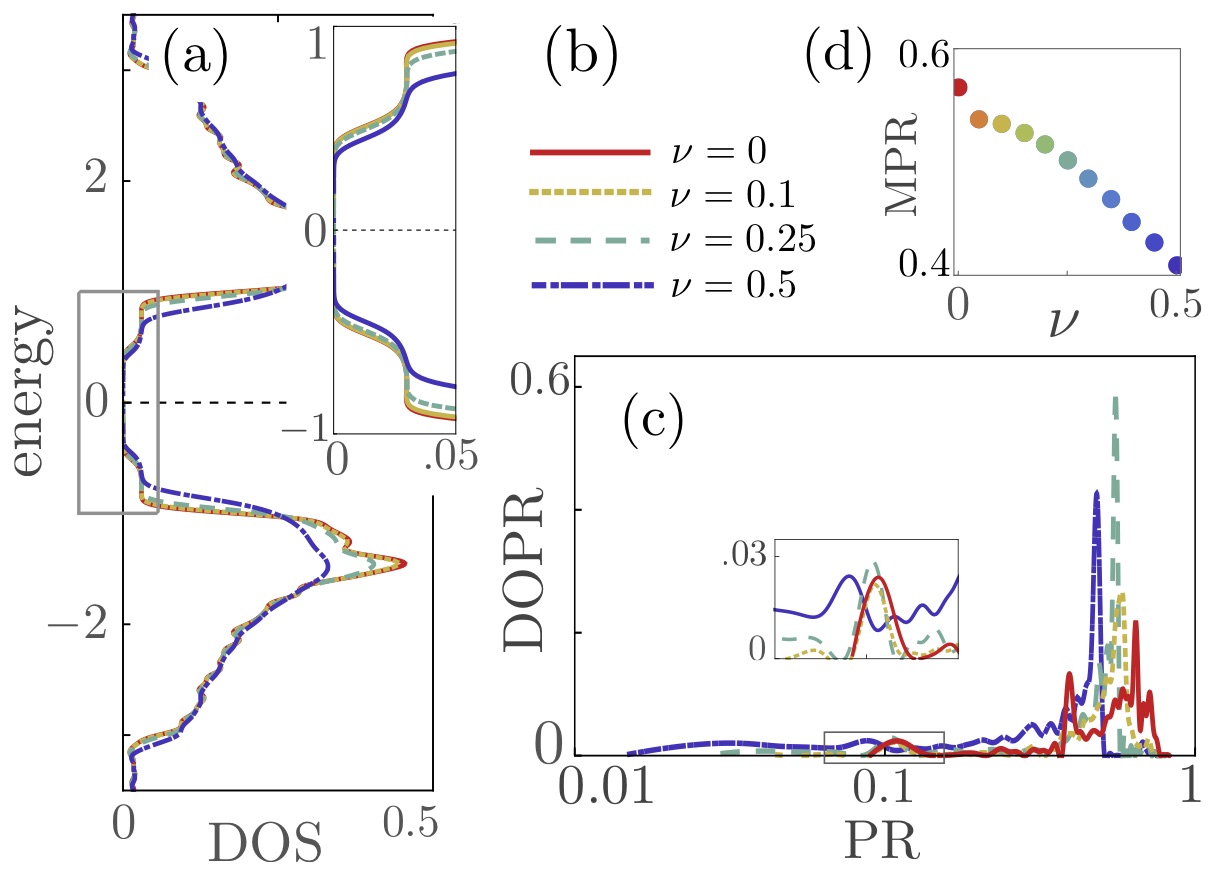}

\caption{This picture illustrates the behavior of a nonlocal-massive Dirac state 
(precisely for $\mu=1$ and $\alpha=1.6$, for a system size $N=1681$)
in the presence of disorder. 
Panel (a) shows the DOS while the inset
is a zoom in to the ingap states. 
Panel (b) is the legend which holds true to all other panels.
Panel (c) shows the DOPR as function of PR for different disorder
strengths, in which the inset gives a zoom in to the peak coming from the edge states.
Panel (d) shows the MPR for a range of disorder strength.}

\label{fig4}
\end{figure}

Fig.\ref{fig4} analyse the results for a representative point within the
phase $\cal D$ (namely $\mu=1$, $\alpha=1.6$, and system size $N=1681$).
Fig.\ref{fig4}a shows the DOS for different disorder strengths. 
First, we clearly observe how the DOS peak decreases with this disorder. Second, we show that the bulk-gap shrinks faster than the edge-gap. In addition, the plateau formed by the massive Dirac edge states (i.e. the finite energies between the bulk-gap and edge-gap) persists quantitatively the same even for large values of disorder, i.e. when compared to the superconducting gap size, which provides an indication of the robustness of the new massive edge states.

One may also look at the Anderson localization effect from the participation ratio (PR), 
which gives the degree of localization of each state after one disorder realization, such that 
\begin{equation}
\text{PR}\equiv\frac{1}{N}\frac{1}{\sum_{\boldsymbol{r}}{\cal P}_{n}^{2}\left(\boldsymbol{r}\right)}.
\end{equation}
For instance, for a completely delocalized state where all sites are
equally probable to be occupied one finds $\text{PR}=1$, while for
a completely localized state where only one site is probable to be
occupied one finds $\text{PR}=1/N$, which goes to zero at the thermodynamic
limit. Moreover, for an edge state perfectly localized at the boundary, i.e. equally distributed
along the edge sites of the 2D system, one finds $\text{PR}=4/\sqrt{N}$. 

Fig.\ref{fig4}c shows the histogram of the participation ratio (which here we call density of participation ratio, DOPR) with respect to the energy index ($n$) for different strengths $\nu$.
Note that our results consider $100$ disorder realizations, and the results are an average over it. 
Thus, in this figure one easily notice that DOPR is concentrating near to $\text{PR}=1$, instead of $\text{PR}\sim10^{-3}$ for this particular system size, which signals that the bulk states are delocalised. 
In addition, we notice that they continue to be delocalized even for large disorder strength,
i.e. we have considered a maximum disorder of $0.5$ while the bulk-gap is nearly $1.0$ (in units of hopping $t$) and the edge-gap is even smaller.
From the edges states we expect a peak near to $\text{PR}\approx0.1$ for this system
size, since they are not localized at one point but spread all over the boundary. 
Thus the inset shows a zoom to the DOPR near $\text{PR}=0.1$.
The existing peaks are clear and they are shifting towards the left when increasing disorder strength, which reflects a trend of the edge states to be more and more localized along the edges.

The spatial localization over all the states is quantified by the mean participation
ratio (MPR), namely
\begin{equation}
\text{MPR}=\left\langle \frac{1}{2N}\sum_{n=1}^{2N}\text{PR}\right\rangle ,
\end{equation}
where the average $\left\langle \cdots\right\rangle $ is over disorder
realizations. Thus, Fig.\ref{fig4}d shows the decreasing
of MPR, roughly from $0.6$ to $0.4$ with $\nu=0$ to $\nu=0.5$,
respectively. This shows a trend of the whole system to become more localized,
besides still orders of magnitude higher than the completely localized value,
typically $\text{PR}\approx6\times10^{-4}$ for this system size.

\subsection{Spatial distribution of states}
\label{sec: spatial distribution}

\begin{figure}
\includegraphics[width=\columnwidth]{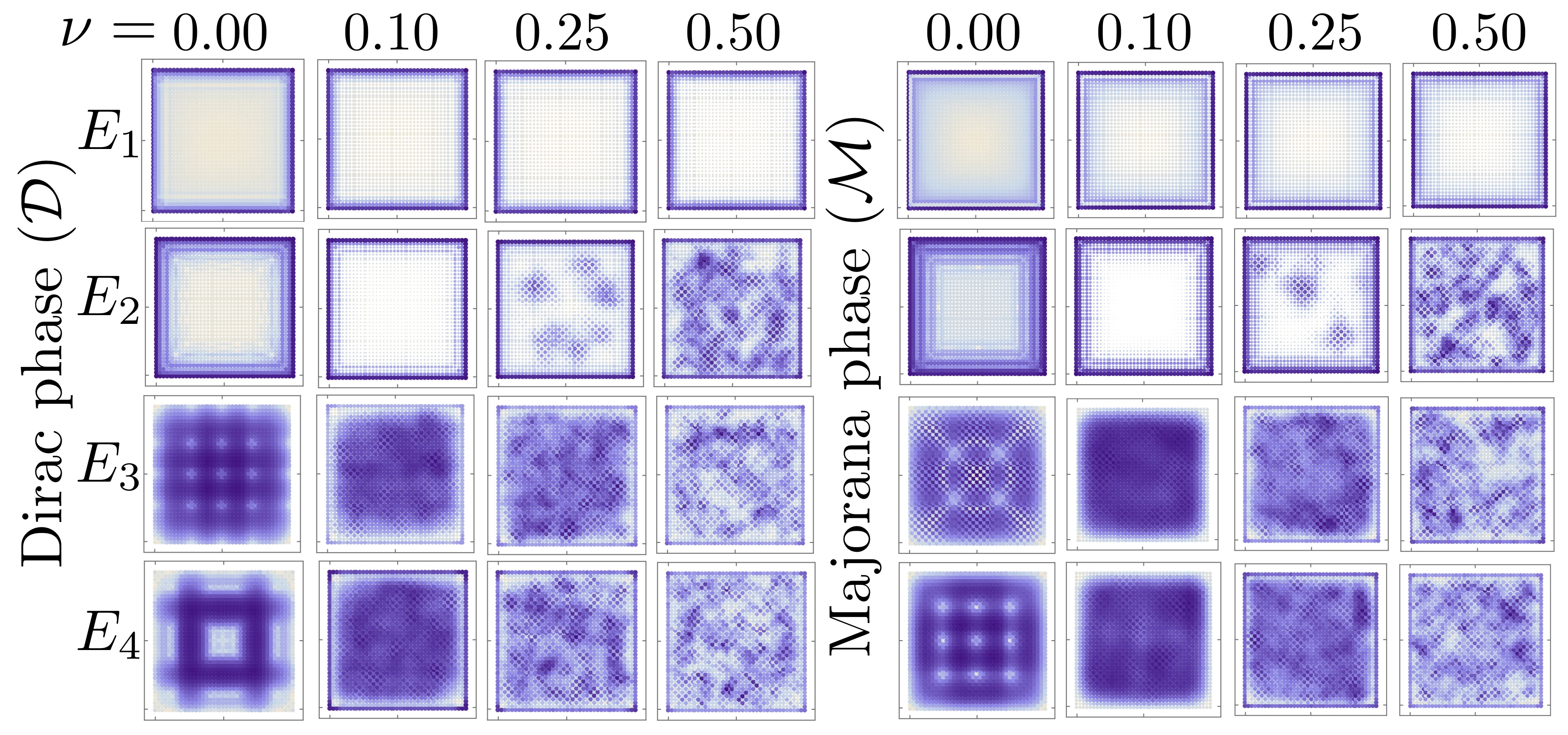}

\caption{Spatial distribution of states within the
phases ${\cal D}$ and ${\cal M}$.
We plotted the probability of occupancy, $\Phi$, associated to the $n$-th energy for a 2D finite-squared system (top view) as described in Fig.\ref{fig1}e-d.
Each row correspond to a different representative quantum state with energy $E_n$, such that: 
$E_1$ is the smallest-finite energy inside the gap;
$E_2$ illustrates the finite energies inside the gap which goes to the bulk with strong enough disorder.
$E_3$ and $E_4$ represent two different bulk energies.
For each of the phases we show what happens to those states after including different values of disorder strength $\nu$.}

\label{fig5}
\end{figure}

Here we analyse the spatial distribution of states subject to static disorder both for the massive Dirac and Majorana phases.
Each row of Fig.\ref{fig5} depicts representative states associated to different energy levels.
We have considered $100$ disorder realizations, and the average was made after sort the energy spectrum and take equivalent energy levels, 
for instance the minimum energy, labelled as $E_1$, was computed as $E_1 = \left< \text{min}(E_{n}) \right>$, where $\left< \cdots \right>$ is the average over disorder realizations and $\text{min}(E_{n})$ takes the minimum energy value among all the energy levels.
The columns of the plot stand for different disorder strengths. 
We notice that the energies $E_1$ to $E_4$ are not the four smallest energies from the energy spectrum, but rather energies which correspond to the following behaviors:
$E_1$ is the smallest finite energy inside the gap;
$E_2$ is a finite energy inside the gap, and will merge to the bulk after including enough disorder.
$E_3$ and $E_4$ are two different energies inside the bulk.

Remarkably, the topological robustness of the massive Dirac phase is indeed very similar to the Majorana phase.
The topological energy states inside the gap display clear localization along the edges with a short tail towards the bulk. We have checked that the tail is shortened after including disorder, adding some degree of additional stability to the boundary of the system.
The increase in edge localization through disorder was already noticed in the inset of Fig.\ref{fig4}c where the peak moves to the left (i.e. towards more localized).
Moreover, Fig.\ref{fig4}a shows that the bulk-gap is shrinking faster than the edge-gap, which means that edge states with higher energies are merging with the bulk. 
This behavior is illustrated in Fig.\ref{fig5} by the frames with energy $E_2$, in which more localised states (like clusters of probability density) are formed inside the bulk. One may notice the formation of those clusters for $\nu \geq 0.25$. 
Finally, the bulk states ($E_3$ and $E_4$) remain fairly delocalized after incorporating disorder. However, for strong disorder we notice the formation of clusters of probability density inside the bulk.

\subsection{Long range disorder}
\label{sec: long range disorder}

\begin{figure}
\includegraphics[width=0.9\columnwidth]{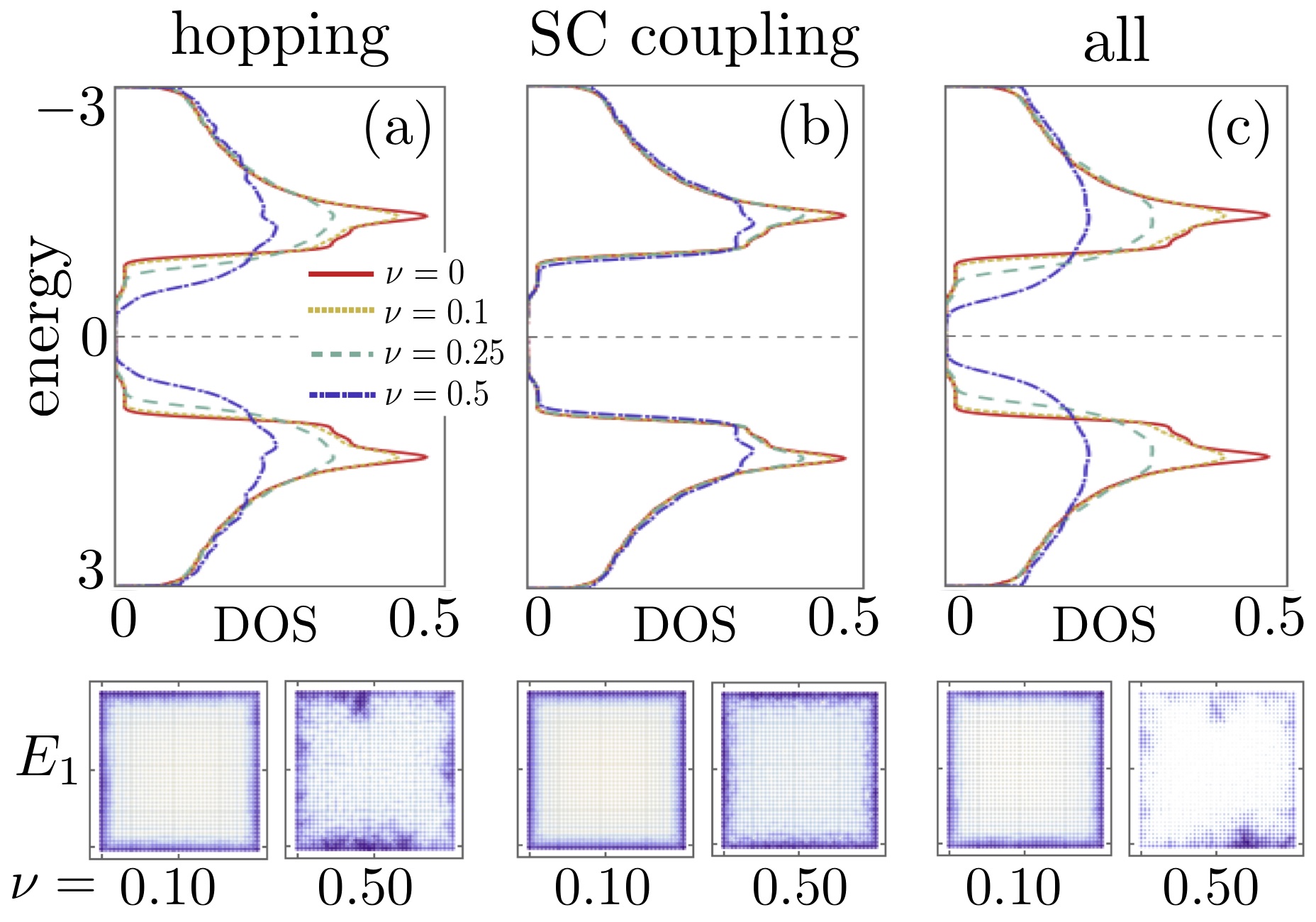}

\caption{Panels (a)-(c) show the DOS in the $\cal{D}$ phase and illustrate the effect of different types of long-range disorder, where we used the same parameters as described in Fig. \ref{fig4}. Panel (a) considers disorder on the hopping strengths, (b) considers disorder on the superconducting coupling strengths, (c) considers both previous cases plus chemical potential disorder.
In the second row we plotted the probability of occupancy, $\Phi$, associated to the $n$-th energy for a 2D finite-squared system (top view) as described in Fig.\ref{fig1}e-d.
In each case, we show the smallest-finite energy inside the gap, $E_1$, for two different values of disorder strength, namely $\nu = 0.10$ and $\nu = 0.50$.}

\label{fig6}
\end{figure}

Some experimental realizations of topological superconductors with long-range couplings may also introduce disorder in the hopping and pairing terms. Therefore, in order to complete the stability analysis of the topological phase, we also introduce disorder perturbations in the hopping and superconducting coupling strengths, and compare their relative robustness. 

The disorder is introduced by replacing $ t  \rightarrow t  +\nu D_{\boldsymbol{r}}\left(R\right)$ and $\Delta\rightarrow\Delta+\nu D_{\boldsymbol{r}}\left(R\right)$ in Eq. (\ref{eq:Hamiltonian real space}), with $\nu$ setting the disorder strength and $\left|D_{\boldsymbol{r}}\left(R\right)\right|\leq1$ is a random number equally distributed over the sites positions $\boldsymbol{r}$ and long-range parameter $R$. 

In Fig. \ref{fig6} we depict three different situations: 
(a) the disorder is included only in the hopping strength; 
(b) the disorder is considered only in the superconducting coupling strength; 
(c) the disorder is included in all couplings, the hopping, the pairing and the chemical potential. 
From panel (a) we notice that long-range disorder affects the edge states more than short-range disorder, however the massive Dirac edge modes are clearly robust against weak and moderate disorder, i.e. the ingap states are present even for $\nu=0.25$, which is already large compared with the size of the bulk gap.
On the other hand, disorder in the superconducting coupling strength is even less harmful. In panel (b) we see a lowering of the gap's peak with enhancing disorder strength but the bulk-gap is nearly constant. 
When compared with Fig. \ref{fig4}a we see that long-range disorder in the superconducting coupling strength affects the system even less than the chemical potential disorder.
Finally, from panel (c) we see that even after including all possible disorder types the largest contribution comes from the hopping, since panels (a) and (c) are very similar.
On the second row of Fig. \ref{fig6} we depict the fate of the massive Dirac modes after including long-range disorder in each case described above. 
Remarkably, even after including a considerable amount of disorder in all couplings, the edge states are still robust and localized. This is explained by the topological nature of the edge states even with long-range couplings.

\section{Discussions}
\label{sec:IV}

We have studied the robustness and localization properties of nonlocal-massive Dirac fermions that appear as exotic energy quasiparticles in 2D topological superconductors with long-range interactions. Analyzing the density of states (DOS) and the energy spectrum, we identify how these topological subgap states at finite energy remain bound to the edge and propagating even for large static disorder. By means of the ingap states we compute the phase diagram for different chemical potentials and long-range couplings.
The propagating massive Dirac fermion is identified from a subgap in the superconducting
phase. Looking at the probability of occupancy of the energy spectrum, we can clearly identify the localization properties of massive Dirac fermions along the edges of a 2D square lattice.
The robustness of these quasiparticles is tested including chemical potential disorder and long-range disorder. The DOS analysis indicates a strong resistance from the ingap states to disorder, which is confirmed using a participation ratio analysis of all quantum states in the system. 
The massive Dirac modes are surprisingly resistant against weak and moderate disorder in the hopping strength, while is practically insensible to disorder in the superconducting coupling strength.
Remarkably, the stability in the probability of occupation for the edge states shows that the robustness of the massive Dirac fermions are analogous to the Majorana states.

Complementarily, for a semi-infinite-periodic system, we notice that a band twisting
in the band structure is always accompanied by a double peak in the DOS. 
We show that this behavior also appears for purely short-range interactions, 
however, we notice it is an exclusive {feature} of topological phases and can be possibly used as a probe to identify nontrivial topology.
In addition, we show that long-range couplings and small pairing strengths strongly enhance the double peak structure.
This enhancement can be potentially used to experimentally detect topological phases using STM measurements \cite{Nadj_et_al14}.

\section{Acknowledgments}

We thank Pablo San-Jose for providing the MathQ package online, and
Tilen Cadez and Liang Fu for useful discussions. This work was supported by Chinese
Agency NSFC under grant numbers 11750110429 and U1530401, Chinese
Research Center CSRC, Fundaci\'on Ram\'on Areces, and RCC Harvard.

\bibliographystyle{apsrev4-1}
\bibliography{refs}

\appendix

\section{Finite size scaling of ingap states}
\label{app: finite size scaling}

\begin{figure}[th!]
\includegraphics[width=0.95\columnwidth]{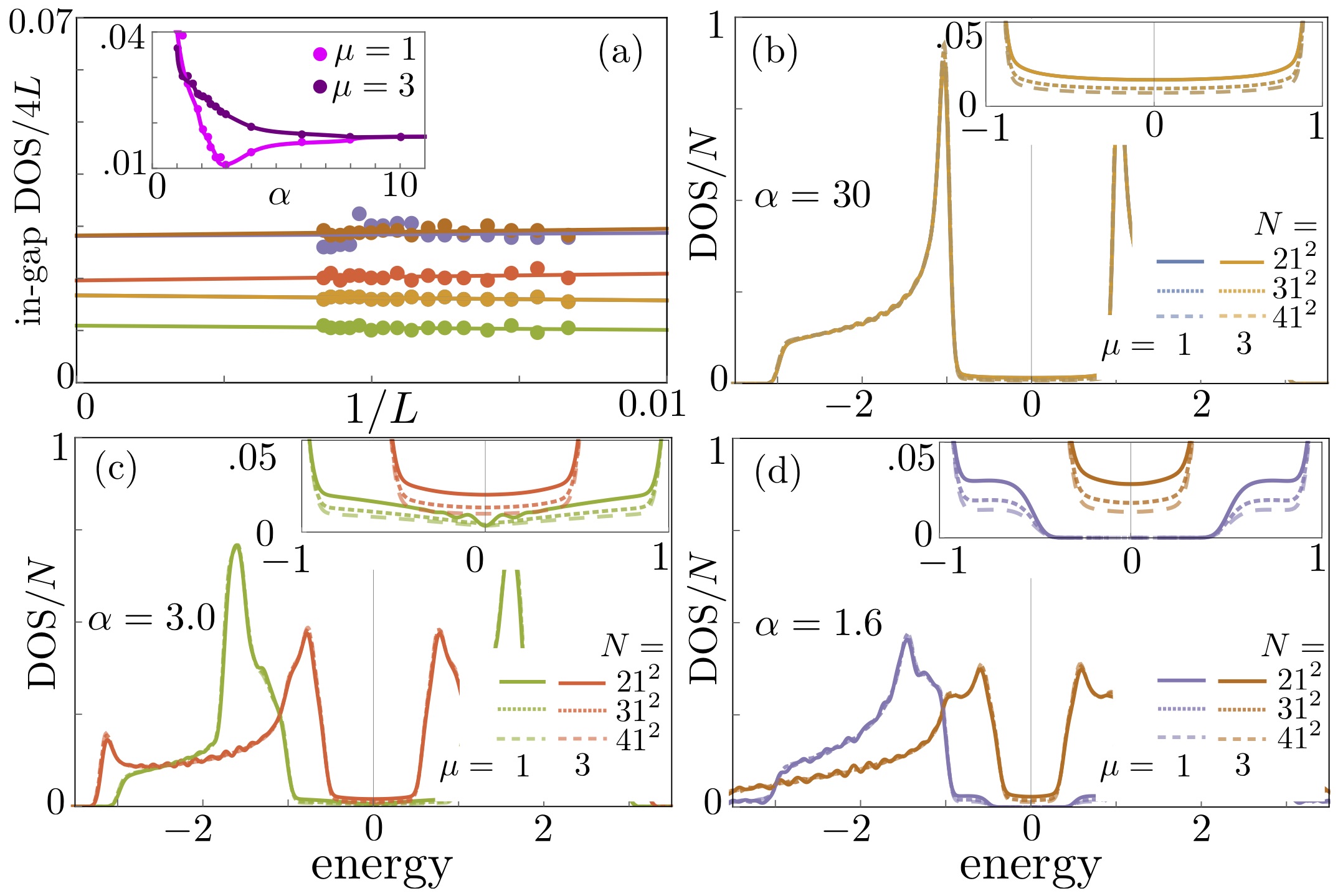}

\caption{Here, we present the finite size scaling analysis of the ingap states.
Panels (b)-(d) show the $\text{DOS}/N$ for three different system
sizes and two representative points in the phase diagram, namely $\mu=1$
and $\mu=3$. Each panel (b)-(d) is computed for a different $\alpha$
value, and the insets are a zoom to the ingap states. 
Panel (a) shows the scaling behavior of the ingap value $\text{DOS}/4L$.
The color code represents different $\mu$ and $\alpha$ and follows the legends of panels (b)-(d),
in particular note that the results corresponding to those parameters used in panel (b) are degenerated.
The inset show the dependance of the ingap states value with the parameter $\alpha$, for the two different values of $\mu$ 
(the solid lines in the inset are guide to the eyes).}

\label{figA1}
\end{figure}

In the main text, Fig.\ref{fig1}c, we show the finite DOS inside
the superconducting gap. Since the bulk states and the ingap states
are expected to have different finite size scalings, here we give
them a detailed analysis. In Fig.\ref{figA1}b-d we show the DOS for
different system sizes and superconducting couplings (controlled by
$\alpha$). In particular, we have used three different system sizes
$N=441,961,1681$, and show results for two representative points
inside the phase diagram, namely $\mu=1$ and $\mu=3$. The insets
show a zoom in to the ingap states. We must notice that here, as well
as in the main text, the DOS is normalized by the system size, i.e.
$\text{DOS}\rightarrow\text{DOS}/N$, which explains why they lay
on top of each other for different system sizes. Thus now we choose
to write this denominator explicitly. On the other hand, the DOS inside
the gap (due to the presence of edges states) are expected to scale
with the perimeter ($4\sqrt{N}\equiv4L$) of the finite system, i.e.
rewriting $\text{DOS}/N\rightarrow\text{DOS}/4L$ one finds ingap
states independent of system size, as shown in Fig.\ref{figA1}a. Finally,
we noticed that the values of the ingap states are dependent on $\alpha$.
The inset in Fig.\ref{figA1}a shows how the exponent $\alpha$ influence the ingap states.
Notice that in the case of $\mu=1$ we have a phase transition, which is accompanied
by the change of $\text{DOS}/4L$ behavior.

\section{Gaussian disorder}
\label{app: Gaussian disorder}

Here we compare two different types of static disorder.
Beyond the random-distributed disorder discussed in the main text, we also make the analysis of Gaussian-distributed disorder, which it is added to the Hamiltonian in Eq. (\ref{eq:Hamiltonian real space}) as
\begin{equation}
H_{\text{disorder}}^{\text{G}}=\nu\sum_{\boldsymbol{r}=1}^{N}D_{\boldsymbol{r}}^{\text{G}}\left(c_{\boldsymbol{r}}^{\dagger}c_{\boldsymbol{r}}-c_{\boldsymbol{r}}c_{\boldsymbol{r}}^{\dagger}\right),
\end{equation}
with $\nu$ setting the disorder strength, and $D_{\boldsymbol{r}}^{G}$ ($\equiv x(\xi)$ in the following) is a random number weighted by the Gaussian distribution with mean value $\mu=0$ and standard deviation $\sigma=0.25$, for each sites' position $\boldsymbol{r}$. Namely, from a random number $\xi$ generated in the range $\xi\in(0,1)$ we can generate a corresponding $x\left(\xi\right)\in(-\infty,+\infty)$ weighted by Gaussian distribution through the equation $x\left(\xi\right)=\mu+\sigma\sqrt{2}\text{err}^{-1}\left(2\xi-1\right)$, where $\text{err}^{-1}$ is the inverse of Error function. This last expression is obtained from the inverse of the cumulant of the Gaussian function. In principle, the cumulant of any normalized distribution can be associated to the random variable $\xi$, in particular for the Gaussian distribution we have $\xi=\int_{-\infty}^{x}\text{e}^{-\left(x'-\mu\right)^{2}/\left(2\sigma^{2}\right)}/\left(\sigma\sqrt{2\pi}\right)dx'$.

\begin{figure}[th!]
\includegraphics[width=0.6\columnwidth]{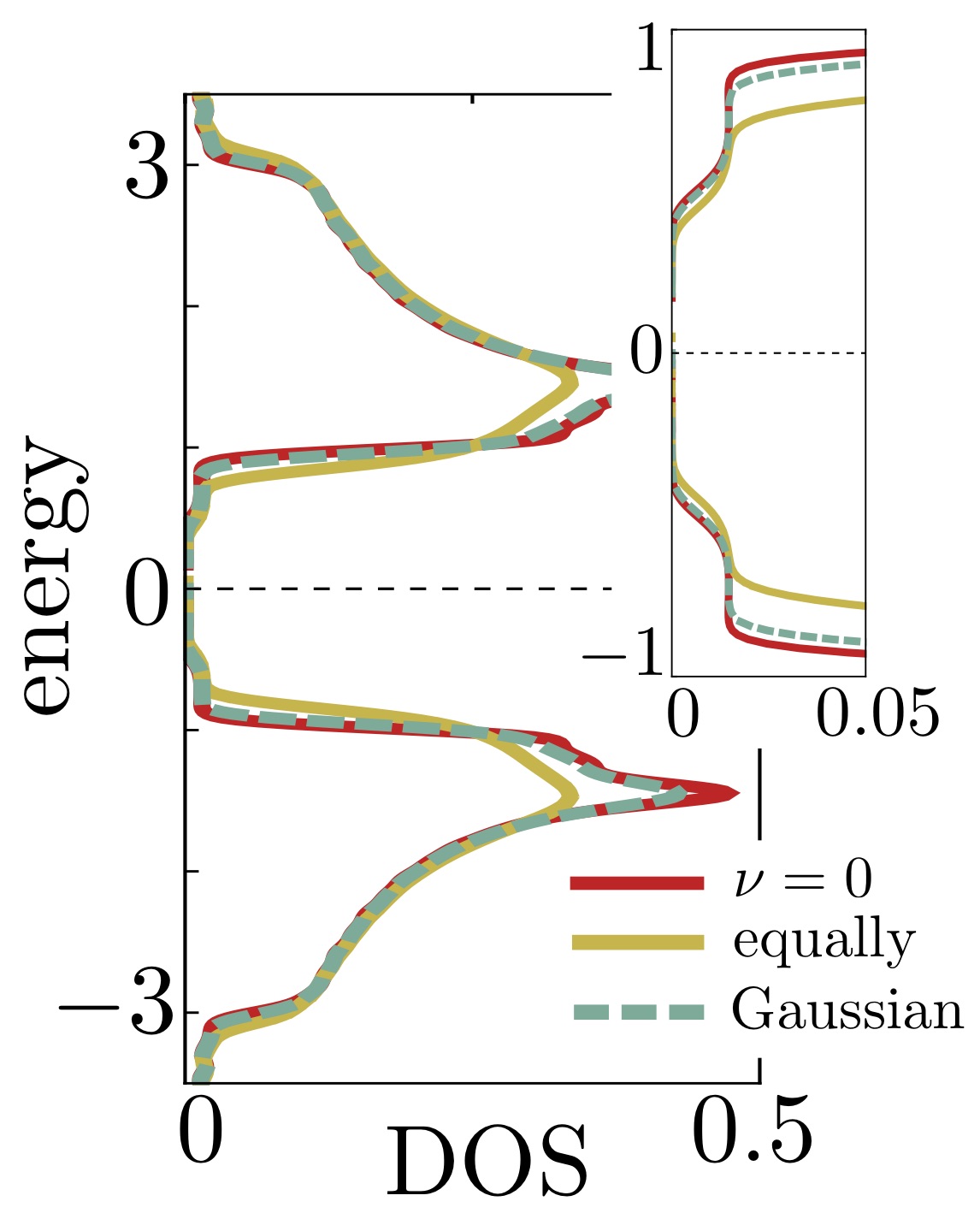}

\caption{This picture illustrates the DOS in the massive Dirac phase (same parameters as in Fig. \ref{fig4}) for different types of static disorder, namely equally-distributed versus Gaussian-distributed disorders in the chemical potential, both computed for $\nu = 0.50$. We also plot the nondisordered case for reference, $\nu = 0$, and the inset shows a zoom to the ingap states.}

\label{figA2}
\end{figure}

As we can see in Fig. \ref{figA2}, Gaussian disorder is less harmful to the system than equally-spaced disorder, which is the case considered throughout the paper as a benchmark for robustness.

\end{document}